\documentstyle[12pt,epsfig]{article}
\def\Journal#1#2#3#4{{#1} {\bf #2} (#4) #3}

\def\PLB{{Phys. Lett.}   {\bf B}}

\def\PRD{{Phys. Rev.}    {\bf D}}

\def\EPC{{Eur. Phys. J.} {\bf C}}

\newcommand {\gapprox}
   {\raisebox{-0.7ex}{$\stackrel {\textstyle>}{\sim}$}}
\newcommand {\lapprox}
   {\raisebox{-0.7ex}{$\stackrel {\textstyle<}{\sim}$}}

\begin{document}
\begin{center}
{\Large\bf Interactions of Heavy Hadrons using Regge Phenomenology
and the Quark Gluon
String Model} \vspace{4mm}\\
Y.R.~de Boer$^{a,b}$, A.B.~Kaidalov$^a$, D.A.~Milstead$^c$, O.I.~Piskounova$^d$\\
\vspace{0.5cm}

 {\it
$^a$Institute for Theoretical and Experimental Physics, Moscow, Russia.\\
$^b$ University of Twente, Enschede, the Netherlands. \\
$^c$ Fysikum, Stockholms Universitet, Stockholm, Sweden.\\
$^d$ P.N.Lebedev Physical Institute of Russian Academy of Science, Moscow, Russia.} \\

\end{center}
 The search for stable heavy exotic hadrons is a promising way to observe new physics processes at collider
experiments. The discovery potential for such particles can be
enhanced or suppressed by their interactions with detector material.
This paper describes a model for the interactions in matter of
stable hadrons containing an exotic quark of charges $\pm
\frac{1}{3}e$ or $\pm \frac{2}{3}e$ using Regge phenomenology and
the Quark Gluon String Model. The influence of such interactions on
searches at the LHC is also discussed.

\section{Introduction}
Searches for slow-moving stable\footnote{The term stable is taken to
mean that the particle will not decay during its traversal of a
detector.} massive particles (SMPs) which interact within a detector
offer a promising means of observing new physics processes at
colliders. Their strong production mechanisms will allow hadronic
SMPs to be among the exotica for which the LHC can open a discovery
window for comparatively small amounts of integrated luminosity
($\sim 1$fb$^{-1}$)~\cite{Fairbairn:2006gg}.
The hierarchy problem suggests the manifestation of hitherto
unobserved physics processes at TeV collision energies, and it is
thus prudent to consider the possibility of heavy exotic stable
quarks. Furthermore, such particles are predictions of
phenomenological implementations of a number of candidate theories
which extend the Standard Model (SM), such as supersymmetry and
universal extra
dimensions~\cite{Fairbairn:2006gg,ff,Gates:1999ei,s5d,sellis,martin}.
An important uncertainty affecting the accuracy of any searches is
the degree to which the detector interactions of hadronic SMPs can
be modelled. In this paper a model is presented for the scattering
in matter of generic heavy hadrons containing either up-like
 or down-like exotic quarks with
charges $\pm \frac{2}{3}e$ and $\pm \frac{1}{3}e$, respectively. The
model is based on Regge phenomenology~\cite{collins} and the Quark
Gluon String Model (QGSM)~\cite{QGSM}.



Collider searches have already ruled out the direct pair production
of a range of SMPs of masses up to around 200
GeV~\cite{Fairbairn:2006gg}. The LHC will open a new discovery
window for SMPs with masses up to several TeV. The ATLAS and CMS
experiments have already developed early SMP search strategies (see,
for example, Refs.
~\cite{Ambrosanio:2000zu,Kraan:2005ji,Johansen:2007kb}), which
usually require a track associated with a slow penetrating particle.
To estimate the efficiency of such a search requires an
understanding of the processes by which a SMP will interact with
detector material. In the case of exotic heavy leptons, only
electromagnetic energy loss could be expected. Hadronic SMPs could,
however, also interact strongly, leading to additional energy loss
and larger rates of SMPs stopped in detector material. Furthermore,
hadronic SMPs could undergo charge exchange reactions, which could
lead to event topologies in which the exotic hadron appears to
possess varying values of electric charge during its passage through
the detector. Models of the interactions which a SMP may undergo are
thus necessary in order to devise effective search strategies which
can not only suppress SM background but also allow SMPs possessing
different quantum numbers to be experimentally distinguished.

While the electromagnetic interactions of massive objects are well
understood~\cite{Yao:2006px}, it is uncertain how hadronic
interactions of an exotic hadron should be treated. Several models
of varying sophistication have been proposed for the scattering of
hadrons which comprise an exotic colour octet, eg
gluino~\cite{Baer:1998pg,mod3,aafke_model,Mackeprang:2006gx}. There
is so far only one detailed
model~\cite{aafke_model,Mackeprang:2006gx} of energy loss and charge
exchange associated with the scattering of heavy hadrons containing
exotic quarks. This approach, implemented within {\sc
Geant}~\cite{geant}, used a black disk approximation to obtain the
total cross section and phase space arguments to predict the
different types of reactions. By using Regge phenomenology and the
QGSM we present in this work a complementary model which can also be
included in {\sc Geant} to aid future searches.


This paper is organised as follows. The mass spectra of hadrons
containing quarks and generic features of their scattering processes
in matter are discussed. The QGSM and Regge-phenomenology are then
used to provide a parameterisation in terms of relevant kinematic
variables of single inclusive particle production arising from
exotic hadron-nucleon scattering. This is used to estimate the
average energy loss in such collisions and the expected rates of
charge and baryon exchange processes.
Using an
ansatz of stable fourth generation quarks, we then describe the impact of
this work on the prospects for detecting exotic hadrons at the LHC.

\section{Properties of exotic hadrons}\label{prop}
When considering interactions with matter, it is important to know
the mass hierarchy of exotic heavy hadrons (hereafter referred to as
$H$-hadrons). This determines the states to which a $H$-hadron, produced
either in the primary interaction or after
scattering with matter, would rapidly decay. Here, we consider only
the lowest lying hadronic states, formed with an exotic heavy colour
triplet charged object $Q$ and light $u$ and $d$
quarks\footnote{Unless stated otherwise, charge conjugate states are
also implied throughout this paper.}.

As outlined in ~\cite{Gates:1999ei,aafke_model}, the lowest lying
neutral and charged mesonic states should be stable since the mass
difference between them is expected to be far smaller than the pion
mass.


The baryon mass spectra, however, are more complicated~\cite{aafke_model}.
The baryon state $Qud$ containing light diquark system of spin 0 is
the lowest lying state $H_{Qud}$ with the other states $H_{Quu}$ and $H_{Qdd}$
forbidden due to the requirement of anti-symmetric baryon wave function.
The heavier states with diquark spin 1 ($H_{Qud}$,
$H_{Qdd}$, and $H_{Quu}$) are, however, all possible. In an analagous
way to the charm sector in which the $\Sigma_c^{0,+,++}$ decays strongly to the $\Lambda_c$
baryon~\cite{Yao:2006px}, it would thus be expected that any produced heavy
baryons would decay to the low mass state $H_{Qud}$ via pion emission. This state
would thus be charged (neutral) for $H$-hadrons consisting of up-like (down-like) exotic quarks. As is shown
in Section~\ref{sec:exp}, this difference in charge of the stable $H$-baryon has a large impact on the
observable rates of $H$-hadrons at the LHC.

In advance of a discovery it is premature to attempt to calculate
precisely the rate and mass of each species of exotic hadron. Here,
it is assumed that $H$-hadrons are degenerate in mass and that a
sample of stable $H$-hadrons formed in high energy collisions, such
as at the LHC, comprise 90\% mesons, divided equally between charged
and uncharged states, and 10\% baryons.


\section{Interactions of $H$-Hadrons in Matter}\label{cross:sec}
Although phenomenological models of the type presented in this work
are needed to predict some of the fine details of exotic hadron
interactions in matter, it is nevertheless possible to build up a
qualitative picture of the scattering
process~\cite{aafke_model,drees}. Owing to the size of its
wavelength, the heavy exotic quark will be a spectator, and it is
the low energy light quark system which interacts.
Thus Regge phenomenology and the QGSM ~\cite{QGSM,kaidalov} can be
employed to model the interactions of exotic hadrons in matter.

\begin{figure}[h!]
  \begin{center}
    \includegraphics[width=0.38\textwidth]{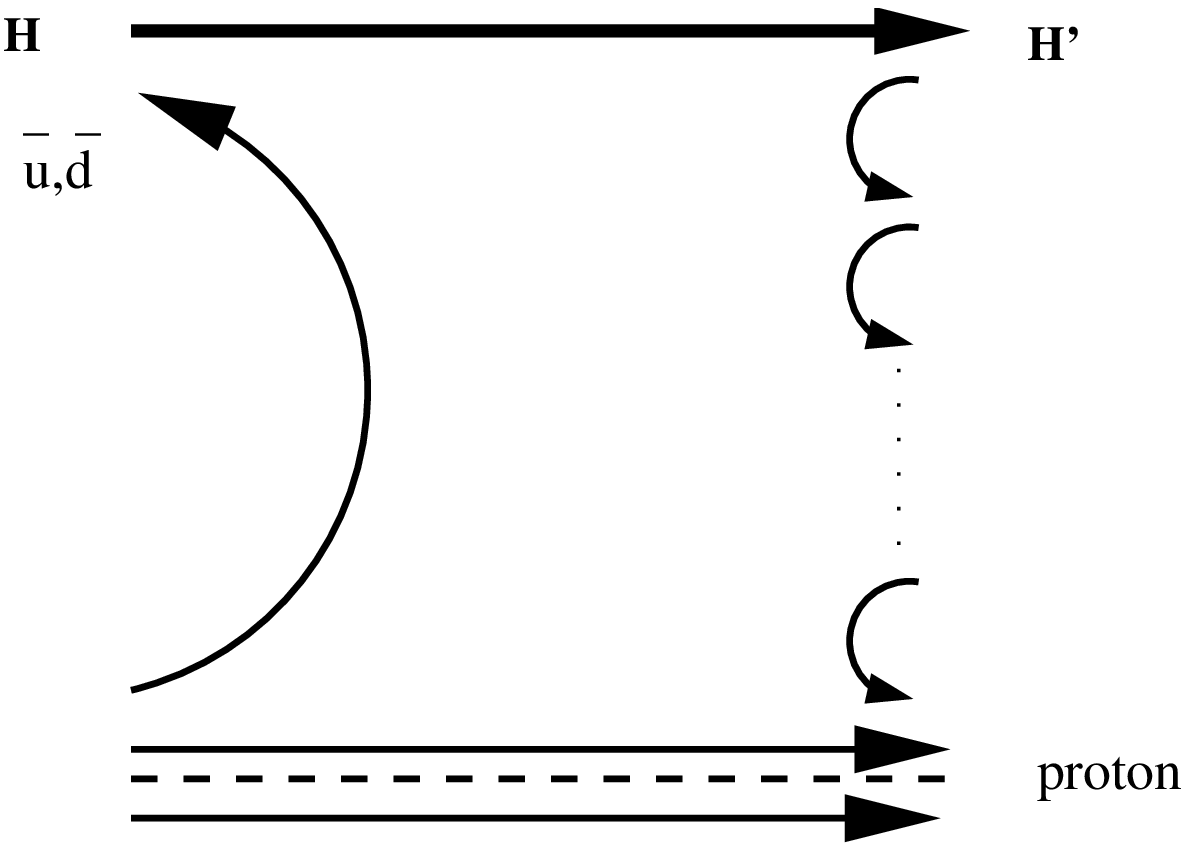}\hspace{1.0cm}
    \includegraphics[width=0.38\textwidth]{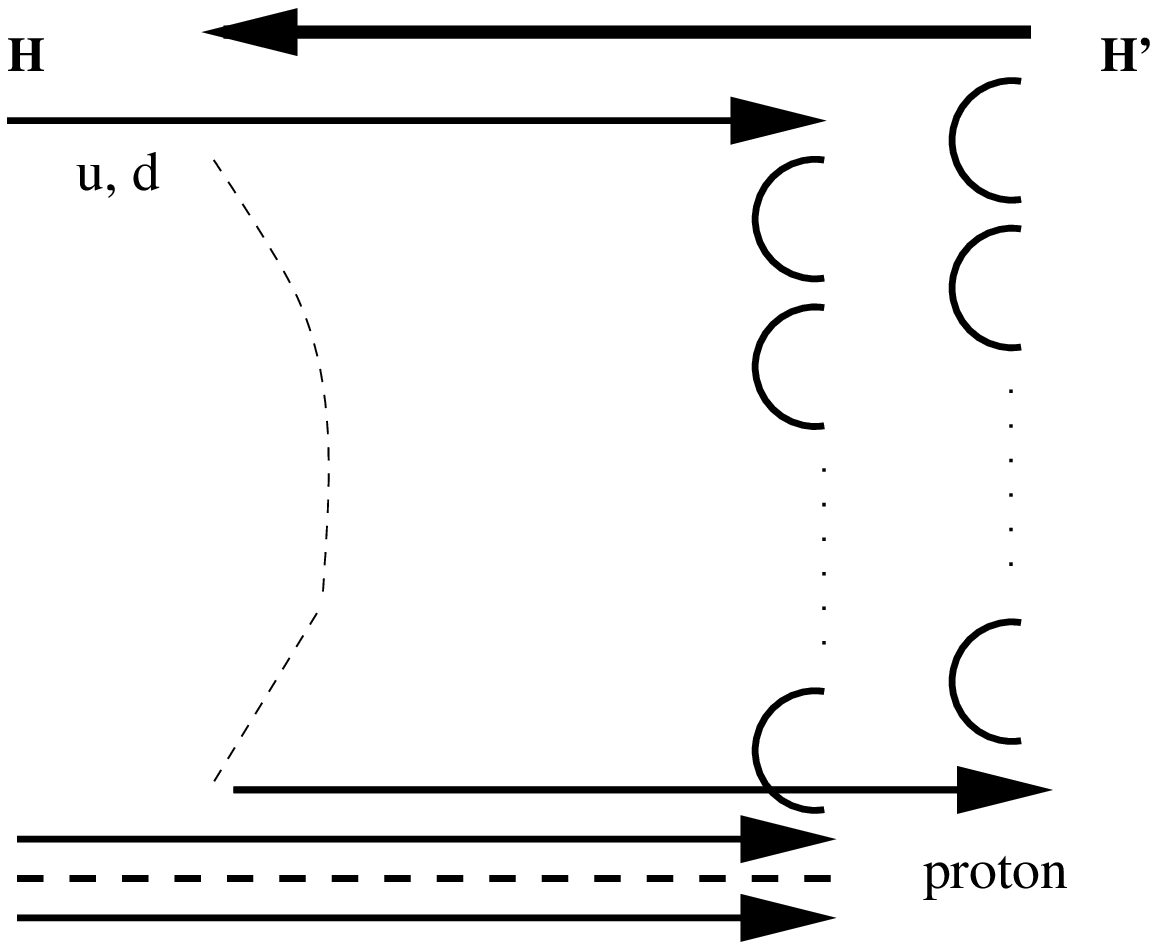}
    \setlength{\unitlength}{\textwidth}
    \caption{Diagrams representing reggeon (left) and pomeron (right) exchange between
a $H$-hadron and a proton.}
    \label{fig_r}
  \end{center}
\end{figure}

The QGSM is based on the $1/N_C$ expansion in QCD
~\cite{Hooft,Veneziano} and partonic interpretation of reggeon
diagrams. All QCD diagrams are classified according to their
topology. In this approach one can distinguish two classes of
scattering processes: reactions mediated by (a) reggeon and
corresponding to planar QCD diagrams and (b) pomeron exchange
related to the cylinder-type diagrams in elastic scattering. Exotic
hadrons containing a light constituent anti-quark, such as a
$H_{Q\bar{q}}$ or a $H_{\bar{Q}\bar{q}\bar{q}}$ interact via
pomeron and reggeon exchanges, the latter processes being due to the
annihilation of light antiquarks with the quarks of detector matter.
Conversely, hadrons containing a light constituent quark
($H_{\bar{Q}q}$, $H_{Qqq}$ ) can only interact via pomeron
exchange. Multiparticle production related to reggeon and pomeron
exchange processes are shown in Fig.~\ref{fig_r}.

\subsection{Total Cross Section of $H$-hadron Scattering}
Let us consider the process of an interaction of a $H$-hadron with a
nucleon of the target nucleus in the target rest frame. In this
frame the light antiquark of a $H$-hadron carries only a small
fraction of the total energy $E$
\begin{equation}
E_q\approx \frac{Em_{q\perp}}{M_H}=\gamma m_{q\perp} \end{equation}
where $\gamma=E/M_H$ and $m_{q\perp}$ is the transverse mass of the
light antiquark. It was shown in the framework of the QGSM
~\cite{QGSM}, that the planar diagram contribution to the total
cross section $\sigma_R(s)$ is universal for the same energy of the
annihilating antiquark. This implies that the contribution to the
total reggeon cross section ($\sigma_R(E)$) can be written as:

\begin{equation}
\sigma_R(E)=K\sigma_{pl}(E=\gamma m_{q\perp})=Kg_R(2\gamma
m_{q\perp}/E_0)^{\alpha_R(0)-1}
\end{equation}
where $K$ is the number of possible planar diagrams and $E_0=1$ GeV.
The vertex parameter $g_R$ can be evaluated from the data on cross
sections of hadronic interactions ~\cite{volkovitski}, and the
intercept of the exchange degenerate Regge tragectories
$\alpha_R(0)$ is equal to 0.5.


The pomeron contribution to the total cross section can be estimated
with two models. In the additive quark model the P-contribution for
the exotic meson is two times smaller than the corresponding value
for pion-nucleon scattering since only one light quark (antiquark)
is present. In the models in which hadrons are considered as colour
dipoles the pomeron cross section is determined by the square of
sizes of colliding hadrons. The mean radius squared $r^2$ of a
hadron containing one heavy and one light quark is about 1/2 of
$r^2$ of a hadron made of two light quarks and we get the same
estimate for P-contribution as for the additive quark model. The
pomeron cross section ($\sigma_P$) depends on energy as
\begin{equation}
\sigma_P\sim (2\gamma m_{q\perp}/E_0)^{\alpha_P(0)-1}
\end{equation} The reggeon contribution to the cross section for a
$H$-meson and a nucleon within a nucleus, which consists of equal
amounts of protons and neutrons, can be derived as the difference
between the reggeon contributions to $\sigma({\pi^-}p)$ and
$\sigma({\pi^+}p)$ data~\cite{collins} multiplied with a factor 1.5.

\begin{figure}[h!]
\centerline{\epsfig{figure=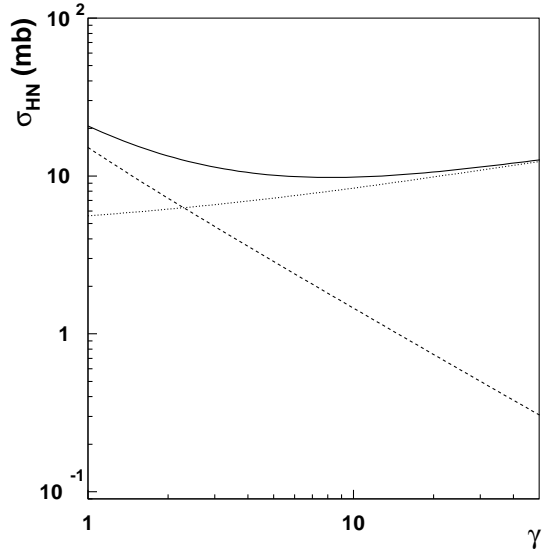,height=8cm,width=8cm}}
\caption{ Pomeron (dashed) and reggeon (dotted) contributions to the
exotic-meson-nucleon cross section. The sum of the two processes is
shown as a solid line.\label{fig_tot}}
\end{figure}

Fig.~\ref{fig_tot} shows the expected cross section for an exotic
meson scattering off a stationary nucleon in a nucleus comprising
equal amounts of protons and neutrons as a function of the Lorentz
factor $\gamma$ of the exotic hadron. The contributions from reggeon
and pomeron exchange processes are shown.

Anti-baryons and baryons may interact via both reggeon and pomeron
exchange, and pomeron exchange only processes, respectively. To
obtain the overall cross sections for interactions involving baryons
and anti-baryons, the pomeron contribution to the meson cross
sections shown in Fig.~\ref{fig_tot} is doubled to take into account
the extra light quark contribution. The reggeon contribution to
anti-baryon scattering is taken to be twice the value for meson
scattering with an added contribution from processes in which exotic
anti-baryons can annihilate to exotic mesons and ordinary mesons. In
the QGSM this process is described by planar diagrams with the
annihilation of diquarks and string-junctions ~\cite{Rossi}. At
large gamma the last process of string-junction annihilation
dominates. In the same way as for usual annihilation it will
decrease with energy as $\sim 1/\gamma^{\frac{1}{2}}$. At
$\gamma\sim 1$ the annihilation cross section can be large ($\sim
30$~mb), and the annihilation cross section is thus taken to be
${30}{\gamma^{-\frac{1}{2}}}$~mb here.

An exotic baryon can convert
 into an exotic meson in one of the sheets of the cylinder diagram of
 Fig.2. This process can however be suppressed at energies close
 to threshold ~\cite{Kraan:2005ji} by phase space effects and the absence of
available pions within the nucleus which would be needed for
reactions in which exotic baryons become mesons.

The Regge approach is valid for $\gamma\gg 1$, however it is known
from experience with hadronic interactions that the Regge description
works reasonably well on average for values of $\gamma\sim
1$~\cite{collins}. The $H$-hadron nucleon scattering cross section
($\sigma_{HN}$) is used to estimate the cross section for the
interaction of a $H$-hadron with a nucleus of atomic number $A$
through $\sigma_{HA}=1.25\sigma_{HN}A^{0.7}$~\cite{geant}.

\begin{figure}[h]
\begin{center}
\epsfig{file=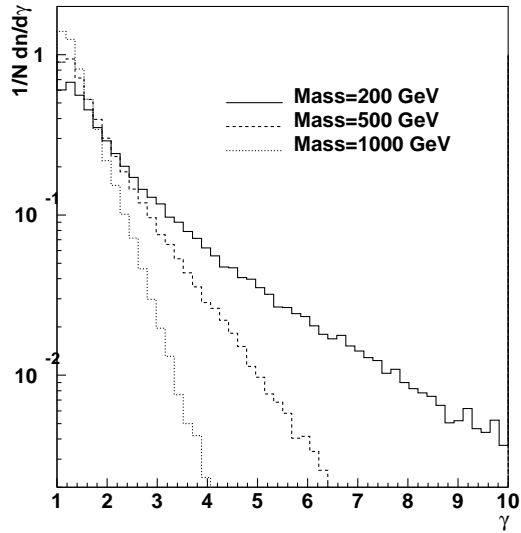,height=8cm,width=8cm} \caption{The
distribution of the Lorentz factor $\gamma$ of fourth generation
quarks pair produced at the LHC, as predicted by {\sc Pythia}. The
spectra are shown for quark masses: 200,
500 and 1000 GeV. \label{fig_gamma} }
\end{center}
\end{figure}

Fig.~\ref{fig_gamma} shows the predicted distribution
$\frac{1}{N}\frac{dn}{d\gamma}$ of the Lorentz factor for fourth
generation quarks pair produced at the LHC, as calculated by {\sc
Pythia}~\cite{pythia}. Here $N$ is the total number of $H$-hadrons
for each mass point: 200, 500, and 1000 GeV. As can be seen, for
increasing mass the quarks will typically be produced with
progressively smaller speeds. It should also be noted that a produced
$H$-hadron must possess a speed above a certain threshold (for ATLAS
this is conservatively estimated as $\gamma \gapprox
1.4$~\cite{Kraan:2005ji}) to satisfy timing requirements
 for the particle to be triggered,
read-out, and assigned to the correct bunch crossing.

\subsection{Differential Cross Sections and Energy losses}\label{sec:res}


In determining the kinematics of the scattering process, we
consider the inclusive process $H+N\rightarrow H' + X$, where $H$,
$H'$, $N$ and $X$ are the incoming exotic hadron, the outgoing
exotic hadron, the target nucleon, and whatever else is produced
in the interaction, respectively. The kinematics of such an
interaction can be specified by three independent kinematic
variables. Commonly used variables are $t$, the usual
four-momentum transfer between the incoming and outgoing exotic
hadrons, $s$, the squared center-of-mass energy of the
interaction, and $M_X$, the mass of the final state $X$.

\begin{figure}[h]
  \begin{center}
    \includegraphics[width=0.38\textwidth]{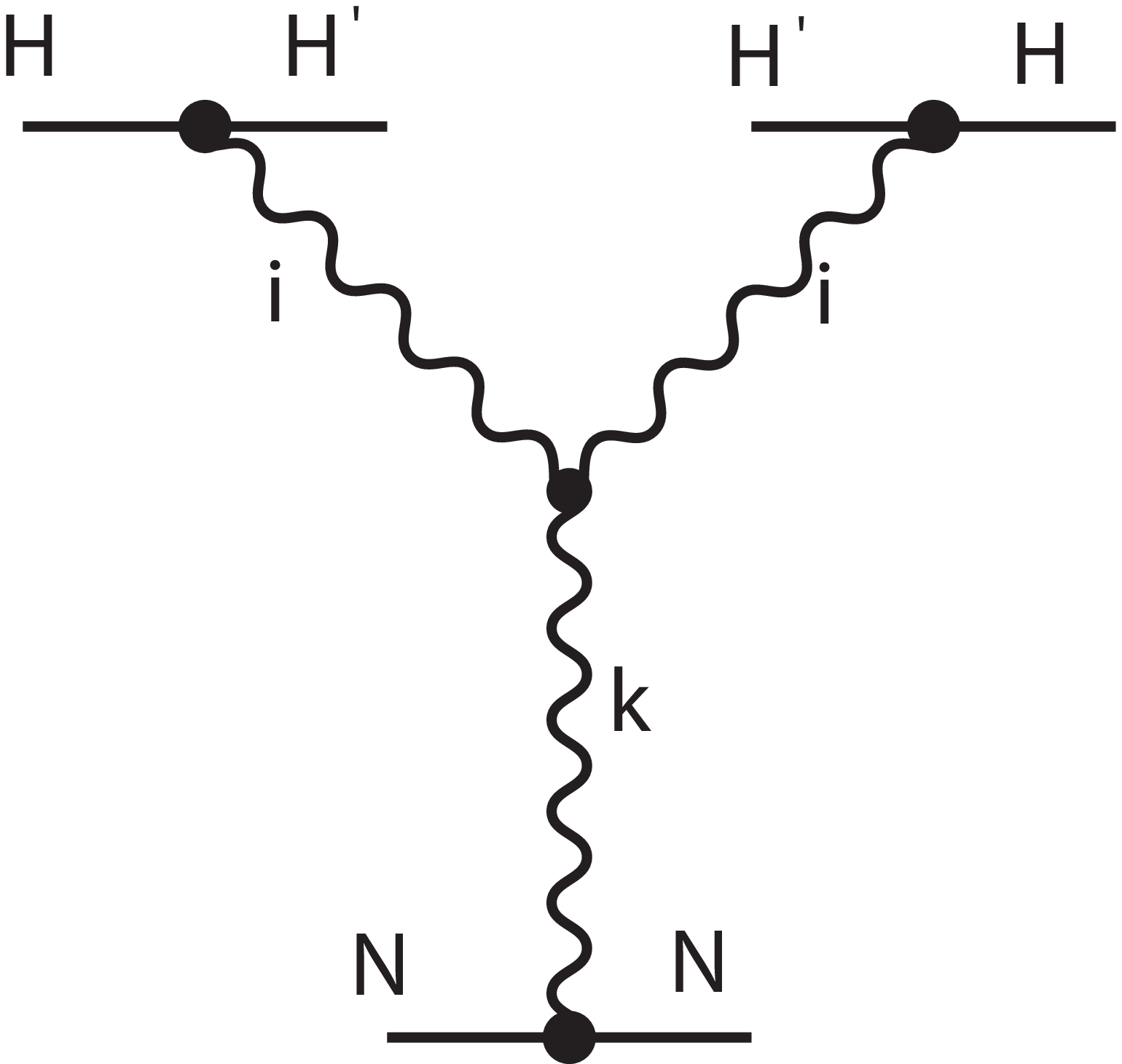}
    \includegraphics[width=0.38\textwidth]{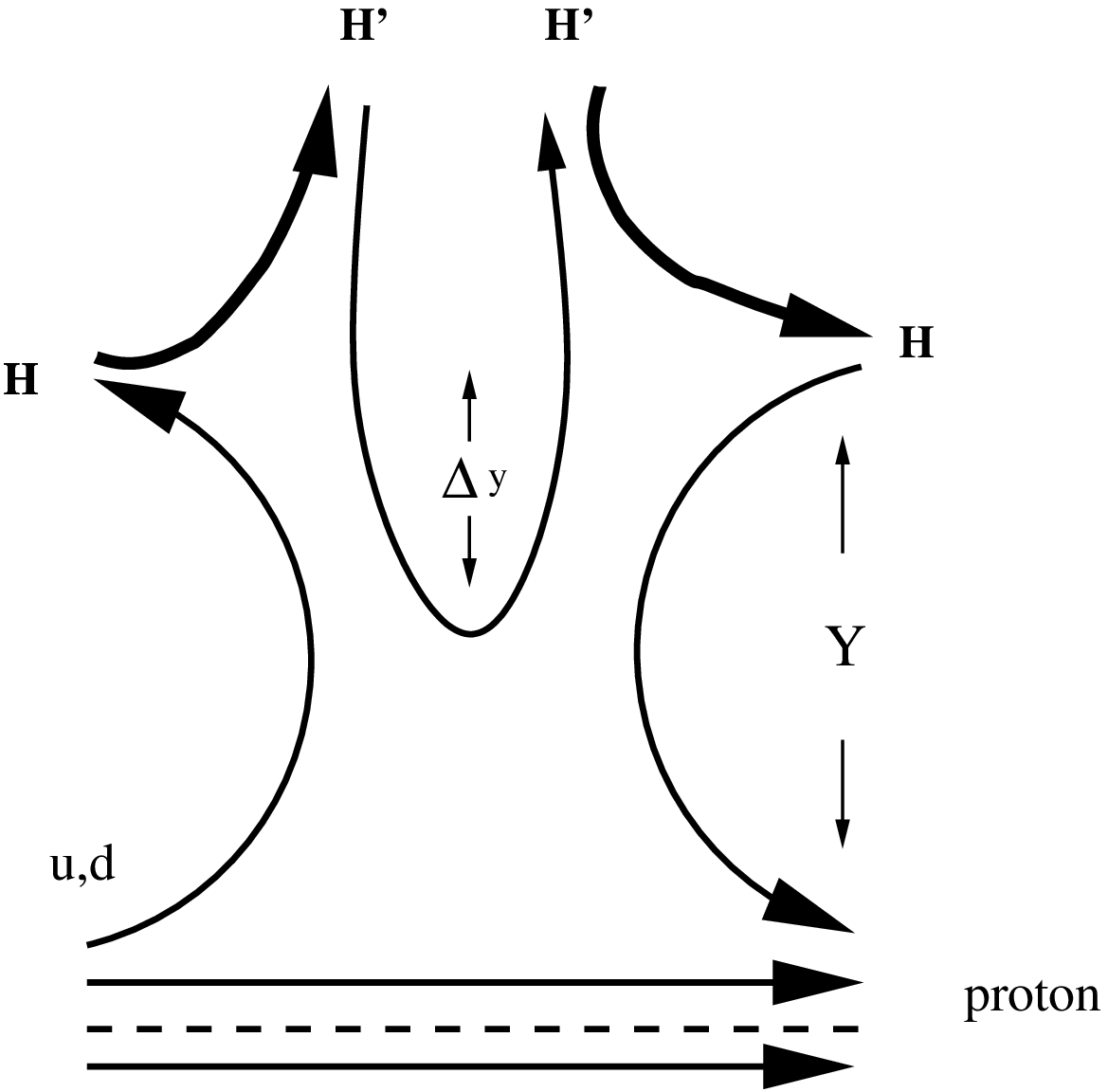}
    \setlength{\unitlength}{\textwidth}
    \caption{Left: triple-regge diagram describing the process $H+N\rightarrow H'+X$.
Right: representation of rapidity gaps between the quark systems
in the three reggeon exchange diagram.}
    \label{fig_tripleregge}
  \end{center}
\end{figure}

The final $H'$-hadron carries a fraction of energy $x_F$ close to
unity and only a small fraction of energy $1-x_F\sim
m_{q\perp}/M_H\ll 1$ is transferred to production of hadrons. This
justifies the application of the triple-regge formulae to provide a
description of inclusive cross sections. Strictly speaking the
triple-regge description is valid for  $m_X^2\gg 1 GeV^2$ and the
rapidity difference between $H'$ and rest hadrons $\Delta y > 1$.
This is equivalent to the condition $2\gamma m_{q\perp}m_N/M_X^2\gg
1$. In hadronic interactions, the triple-regge description works
usually up to $\Delta y \sim 1$ and we will assume in the following
that the same is true for interaction of $H$-hadrons~\cite{collins}.  A representation of
rapidity gaps between the quark systems in the three reggeon exchange diagram is
shown in Fig.~\ref{fig_tripleregge}.

Expressions for the contributions of different triple-regge terms
$iik$ (Fig.~\ref{fig_tripleregge}) to inclusive cross sections is
straightforward to obtain, noting that for reggeons $i$ corresponds
the factor $\exp(2(\alpha_i(t)-1)\Delta y)$, while an exchange by
the reggeon $k$ leads to the factor $\exp((\alpha_k(0)-1)y_q)$.
Here, $y_q=\ln(M_X^2/(m_{q\perp}m_N)$ is the rapidity interval
covered by produced hadrons (the total rapidity
$Y=\ln(2E/M_H)=\ln(2\gamma)=\Delta y + y_q)$. As for the total cross
section we consider as exchanged reggeons $i,k$ the pomeron P and
secondary reggeons R. Thus we have the following triple-regge
contributions: RRR,RRP,PPR and PPP.

Using the rules described above we can write inclusive cross
sections for the corresponding triple-regge terms in the following
forms:
\begin{eqnarray}
\frac{d^2\sigma_{RRR}}{dtdM^2_X}(\gamma,M^2_X)&=&\frac{1}{M^2_X}\sigma^2_R(\gamma)C_{RRR}
\exp[(2B_{RH}+B_{RRR}+2\alpha_R^{\prime}\ln(\frac{2\gamma M^2_0}{M^2_X}))t]\left(\frac{M^2_0}{M^2_X}\right)^{\Delta_R}\\
\frac{d^2\sigma_{RRP}}{dtdM^2_X}(\gamma,M^2_X)&=&\frac{1}{M^2_X}\sigma^2_R(\gamma)C_{RRP}
\exp[(2B_{RH}+B_{RRP}+2\alpha_P^{\prime}\ln(\frac{2\gamma M^2_0}{M^2_X}))t]\left(\frac{M^2_0}{M^2_X}\right)^{2\Delta_R-\Delta_P}\\
\frac{d^2\sigma_{PPR}}{dtdM^2_X}(\gamma,M^2_X)&=&\frac{11}{M^2_X}\sigma^2_P(\gamma)C_{PPR}
\exp[(2B_{PH}+B_{PPR}+2\alpha_P^{\prime}\ln(\frac{2\gamma
M^2_0}{M^2_X}))t]\left(\frac{M^2_0}
{M^2_X}\right)^{2\Delta_P-\Delta_R}\\
\frac{d^2\sigma_{PPP}}{dtdM^2_X}(\gamma,M^2_X)&=&\frac{1}{M^2_X}\sigma^2_P(\gamma)C_{PPP}
\exp[(2B_{PH}+B_{PPP}+2\alpha_P^{\prime}\ln(\frac{2\gamma
M^2_0}{M^2_X}))t]\left(\frac{M^2_0}{M^2_X}\right)^{\Delta_P}
\end{eqnarray}
where $\Delta_R= \alpha_R(0)-1$= -0.5, $\Delta_P=\alpha_P(0)-1$=
0.12, $\alpha_R^{\prime} = 0.9~GeV^{-2}$, $\alpha_P^{\prime} =
0.25~GeV^{-2}$ ~\cite{QGSM} and $M^2_0=m_N m_{q\perp}=0.5~GeV^2$.

The parameters, $C_{iij}$, and $B_{iij}$ can be determined using
Regge factorization from the triple-regge description of inclusive
spectra in high-energy hadronic interactions~\cite{kaidalov}. Let us
emphasise that the RRR-term corresponds to the diagram of
Fig.~\ref{fig_r} (left), which represents the cutting of the planar diagram or R-exchange,
while the RRP-term corresponds to the cutting of the cylinder-type
diagram in Fig.~\ref{fig_r} (right). Due to conservation of $H$-hadrons
integrals over $M_X^2$ and $t$ give $\sigma_R$ and $\sigma_P$
contributions to the total cross section correspondingly.

The PPR and PPP-terms describe the diffractive dissociation of a
nucleon and their cross sections can be calculated, using
factorization from the corresponding cross sections extracted from
$pp$-interactions
\begin{equation}
\sigma^{PPi}_{Hp}=\frac{\sigma_P(Hp)^2}{\sigma_P(pp)^2}\sigma^{PPi}_{pp}
\end{equation}
Here, we neglected the small difference in $t$-dependence for $Hp$
and $PP$ vertices. Taking into account that
$\frac{\sigma_P(Hp)}{\sigma_P(pp)}\approx 1/4$ and that the sum of
PPR and PPP-contributions for $pp$-collisions in the relevant energy
domain does not exceed 2mb, we obtain very small cross sections for
diffraction dissociation of a nucleon in $Hp$-interactions: 0.12 mb.
Thus these cross sections constitute only about 1$\%$ of the total
cross section and can be safely neglected in our estimates of energy
losses.

For parameters characterising the $t$-dependence of RRR and
RRP-terms we take the same values as have been extracted from
analysis of $pp$-interactions ~\cite{kaidalov}:
  $2B_{RH}+B_{RRR}=2B_{RH}+B_{RRP}= 4~GeV^{-2}$

The energy loss of a $H$-hadron is given by:
\begin{equation}
\Delta{E}=\frac{M_X^2-m_N^2 + |t|}{2m_N}
\end{equation}
The average energy loss can thus be calculated:

\begin{equation}
\label{avde}
 \langle E \rangle = \frac{
   \int_{m_{N}+m_{\pi}}^{M_{Xmax}} dM_{X} \int_{|t|_{min}}^{|t|_{max}} d|t|
   \Delta E \frac{d^{2}\sigma}{d|t|dM_{X}}
 } {
   \int_{m_{N}+m_{\pi}}^{M_{Xmax}} dM_{X} \int_{|t|_{min}}^{|t|_{max}} d|t|
   \frac{d^{2}\sigma}{d|t|dM_{X}}
 }
\end{equation}

Here, $m_{N}$ and $m_{\pi}$ were taken as the mass of the proton and
a charge pion, respectively.

The upper limit on $M_X$ is taken to be the lower of the following
two limits: $M_{Xmax}=(2\gamma M^2_0)^\frac{1}{2}$, which represents
the condition $\Delta y = 0$ or $M_{Xmax}=\sqrt{s}-m_{H}$ from
energy-momentum conservation and where $m_{H}$ is the mass of the
interacting $H$-hadron.

The limits on $t$ are given by
\begin{equation}
 |t|_{min,max}(M_X)=2[E(m_N)E(M_X)\mp
p(m_n)p(M_X)-m_H^2]
\end{equation}

 where
$E(m)=\frac{s+m_H^2-m^2}{2\sqrt{s}}$,
$p(m)=\frac{\lambda^{\frac{1}{2}}(s,m_H^2,m^2)}{2\sqrt{s}}$, and
$\lambda(a,b,c)=a^2+b^2+c^2 -2(ab+ac+bc)$.

\begin{figure}[h!]
\begin{center}
\epsfig{file=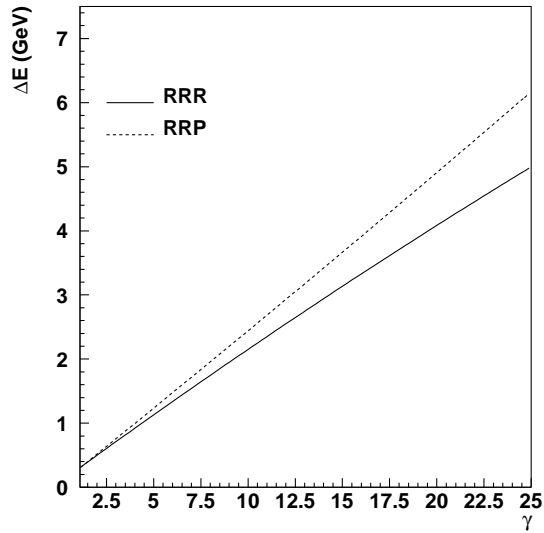,height=8cm,width=8cm} \caption{Average
energy loss per interaction as a function of $\gamma$ for two triple-regge
terms.\label{fig_et} }
\end{center}
\end{figure}

Fig.~\ref{fig_et} shows the mean energy loss associated with two
triple-regge contributions (RRR, RRP) as a function of $\gamma$. The
two contributions give similar values except at the very highest
values of $\gamma$, at which the production cross section for
$H$-hadrons is tiny. It was also found that the energy loss at a
fixed $\beta$ is insensitive to the $H$-hadron mass for mass values
greater than around 10 GeV. The principal uncertainty in the total energy loss comes from the
uncertainty in the value of $m_{q\perp}$, which enters via the parameter
$M^2_0$ and was chosen to be 0.5~$GeV^2$ for this work. Taking
values of $M^2_0$ of 0.3 or 0.7 GeV$^2$ changes the average energy
values typically by around 50\% and this would not change the
conclusions of this paper.

 For
calculations of energy losses of a given $H$-hadron type we need in
principle to know only the losses associated with  each triple-regge
term and the relative weights of different terms. For RRR and
RRP-terms these weights are given by
$\sigma_R(\gamma)/\sigma^{tot}(\gamma)$ and
$\sigma_P(\gamma)/\sigma^{tot}(\gamma)$ correspondingly. However,
since  the difference in energy loss from the two contributions is
anyway very small,  $H$-hadron energy loss for both reggeon and
pomeron interactions was assumed to be that given by the RRP term.

\begin{figure}[h!]
\begin{center}
\epsfig{file=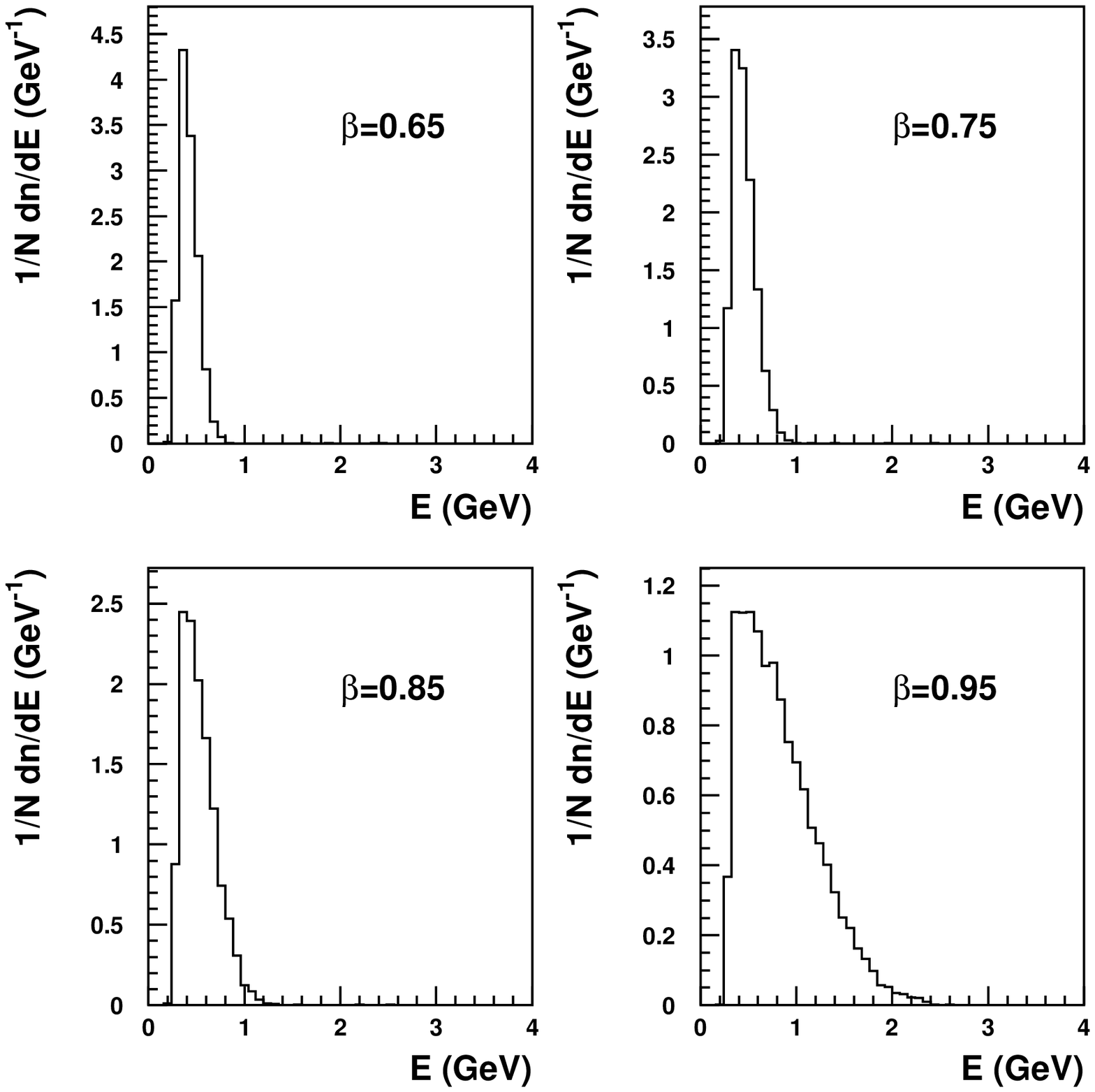,height=10cm,width=10cm} \caption{The predicted
distribution $\frac{1}{N}\frac{dn}{dE}$ of the energy loss per interaction of a
$H$-hadron of mass 200 GeV and for different values of $\beta$.
\label{fig_enloss} }
\end{center}
\end{figure}

A Monte Carlo method was used to extract the predicted distributions
$f(E)=\frac{1}{N}\frac{dn}{dE}$ of energy loss ($E$) at a range of
values of $\beta$, where $N$ is the total number of $H$-hadrons at
each $\beta$ point. The functional form $f(E)$ for RRR and RRP terms
in the $H$-hadron $\gamma$ range expected at the LHC ($\gamma
\lapprox 10$) can be approximated by Equation~\ref{eqf}.
\begin{equation}
f(E)=\exp[{-10(E-0.3)^2(1-\beta)^{1.5}}]\frac{1.0}{1+(\frac{0.3}{E})^{15}}
\label{eqf}\end{equation}
This function was used to arrive at the
results presented in Section~\ref{sec:exp}. Fig.~\ref{fig_enloss}
shows $f(E)$ for an exotic $H$-hadron of mass 200 GeV for four
values of $\beta$. The distribution typically extends up to around a
few GeV, though it is peaked at around $0.5$ GeV.


\subsection{Charge Exchange and Baryon Formation}
Charge exchange processes will naturally take place via the
formation of light quark pairs from the vacuum both for planar
(Fig.1) and cylinder-type (Fig.2) diagrams. As $u$ and $d$ quarks
are produced with equal probability we expect that charge exchange
happens with 50$\%$ probability per interaction.

Within the Lund string model~\cite{string} exotic baryons only
account for around $10\%$ of all exotic hadrons produced at the
primary interaction, a prediction which is also made in the
QGSM~\cite{olga2}. Baryons can also be produced as a result of
hadronic interactions of exotic mesons in matter. The probability
that a given inelastic collision involving an exotic meson results
in baryon formation takes place is taken to be 10$\%$, a value
motivated by investigation of baryon production in QGSM and low
energy hadron scattering data~\cite{QGSM}. The proportion of
different baryon species formed in such interactions is taken to be
the same as at the primary interaction.

\section{Experimental Signatures}\label{sec:exp}
In this section the model of hadronic scattering outlined in
Section~\ref{cross:sec}, together with the well established theory
of electromagnetic energy loss of charged particles in
matter~\cite{Yao:2006px}, is used to estimate the energy losses
and rate of charge exchange reactions of a $H$-hadron propagating
through the ATLAS detector~\cite{tdr} at the LHC. Calculations of
the cross section for fourth generation quark pair production at the
LHC are then used to predict the rates of various event topologies
associated with $H$-hadron production. The event topologies are characterised by
the values of charges possessed by the $H$-hadrons in the inner and muon
 tracking chambers, i.e. before and after scattering in the ATLAS calorimeter system.


\subsection{Energy Loss and Charge Exchange}\label{eloss}
 To estimate the kinematic distributions of
$H$-hadrons produced at the LHC, the {\sc Pythia}~\cite{pythia}
program was used to generate samples of 50,000 fourth generation
quark pair production events for quark masses 200, 500 and 1000 GeV.
It was assumed that the proportion of different types of stable
$H$-hadrons formed from these quarks follows the prescription given
in Section~\ref{prop}. To ensure that $H$-hadrons would belong to a
high acceptance region of the ATLAS detector, the initial value of
$\beta$ of the $H$-hadrons was required to be greater than 0.7 and
the pseudorapidity was restricted to
$|\eta|<2.5$~\cite{Kraan:2005ji}. The remaining $H$-hadrons were
then transported, using a Monte Carlo method, through material of
thickness corresponding to the part of the ATLAS detector enclosed
by the muon tracking chamber.

The material distribution of the sub-systems of the ATLAS detector
enclosed by the muon detector varies between
around 11 and 19 interaction lengths as a function of pseudorapidity~\cite{tdr}.
The largest material contribution ($\sim 70\%$ of the total thickness) arises
from the absorbing material in the different hadronic calorimeter
systems. These systems are the Tile Calorimeter
(TileCal)~\cite{tile} and Hadronic EndCap Calorimeter
(HEC)~\cite{endc}, which use iron and copper, respectively, as
absorbers. Thus, for a generated $H$-hadron falling in the
pseudorapidity regions $|\eta|<1.5$ ($1.5<|\eta|<2.5$) covered by
the TileCal (HEC), iron (copper) was used to represent the ATLAS
detector material in the Monte Carlo calculations presented here.

\begin{figure}[h!]
\begin{center}
\epsfig{file=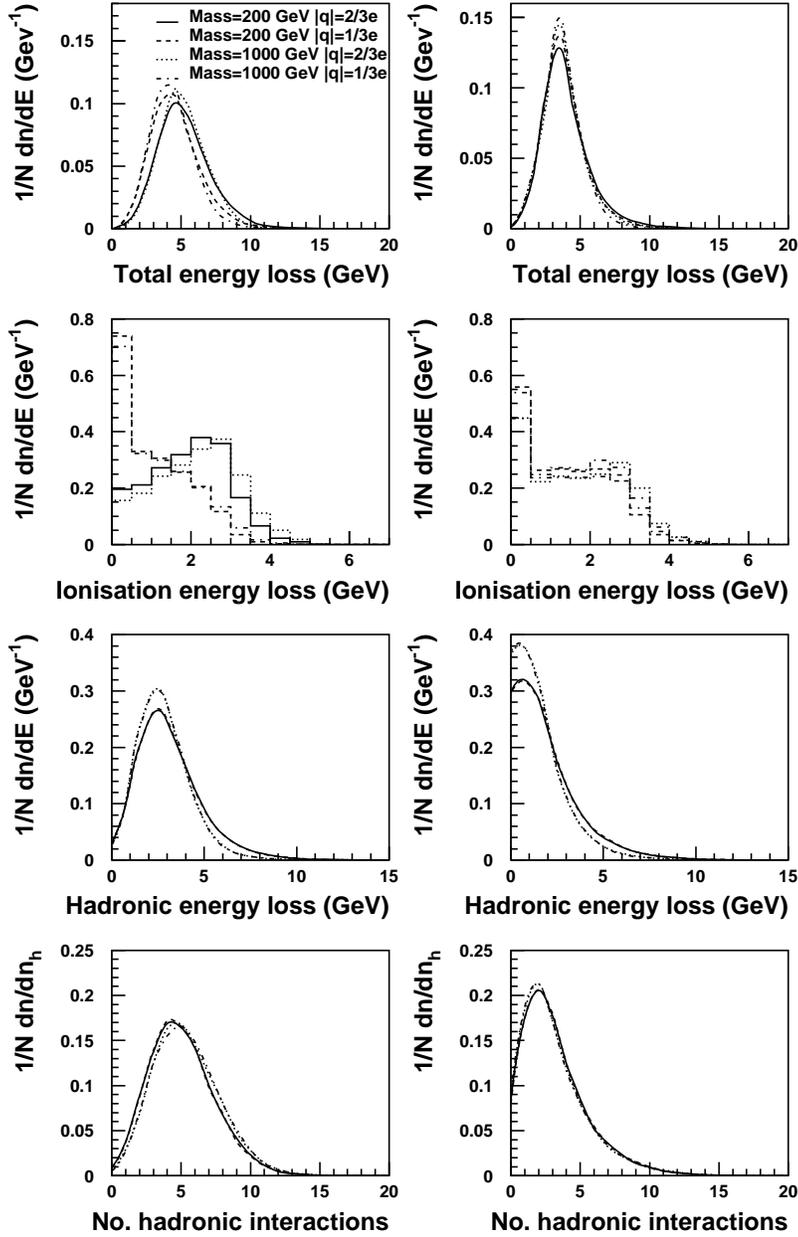,height=18cm,width=12cm}
\caption{Distributions of energy loss and hadronic scattering for
$H$-hadrons of masses 200 and 1000 GeV and for exotic quarks of
charges $\pm \frac{1}{3}e$ and $\pm \frac{2}{3}e$. The left (right) column represents $H$-hadrons containing
an exotic quark (anti-quark). Distributions of the total, ionisation and hadronic energy loss is shown
 along with the multiplicity of interactions. The distributions assume no mixing of neutral
$H$-mesons. \label{fig_cont1}}
\end{center}
\end{figure}

Fig.~\ref{fig_cont1} shows differential distributions related to the
energy loss and interactions of $H$-hadrons as they pass through the
detector material. Spectra are shown separately for $H$-hadrons formed from
different types of exotic quarks and anti-quarks with masses 200 and 1000
GeV. The distributions are normalised to the total number $N$ of a given type of
$H$-hadron satisfying the $\beta$ and $\eta$ requirements. In the following discussion of the plots
$H_{Q} (H_{\bar{Q}}$) is used as a generic
term to denote a $H$-hadron with an exotic quark (anti-quark) while the terms
$H_{U},H_{D},H_{\bar{U}}$, and $H_{\bar{D}}$ are used to denote $H$-hadrons with up-like and down-like
exotic quarks and anti-quarks.

The distributions of total energy loss are peaked at $\sim$4-5~GeV
and extend up to around 10~GeV for the different $H$-hadron types,
with no substantial mass dependence. The $H_{D}$-hadrons typically
show lower energies than the $H_{U}$-hadrons. This is due to
differences in ionisation energy loss since $H_{D}$-hadrons are more
likely to
 propagate through the material with zero electric charge. A $H_{D}$-hadron can, for example, start as a neutral
meson and then be converted into a neutral baryon. The effect can
be seen as a peak at low ionisation energy low for $H_{D}$-hadrons. Smaller peaks at
 low ionisation energy loss
are also
visible for the $H_{\bar{Q}}$-hadrons, which arise from events in which those $H$-hadrons
propagate through the material mostly as neutral mesons.

The hadronic energy
 loss for the $H$-hadrons
decreases slightly with mass owing to the typically lower speeds of the more massive $H$-hadrons
 (see Fig.~\ref{fig_et}). The hadronic energy loss for
 $H_{Q}$-hadrons peaks at around 3-4~GeV, which is larger than that for
 $H_{\bar{Q}}$-hadrons ($\sim 1$~GeV). This can be understood as a consequence of the different reaction channels, which
are open for $H_{Q}$ and $H_{\bar{Q}}$-hadrons. $H$-hadrons are
  dominantly mesonic following hadronisation and, although they can form baryons, still
 interact mostly as mesons. As explained in Section~\ref{cross:sec}, $H_{Q}$-mesons  may scatter
 via reggeon or pomeron exchange, unlike $H_{\bar{Q}}$-mesons, which interact with a
 lower cross-section since only pomeron exchange is possible in this case. This is seen in the distribution
 of the multiplicity of hadronic interactions, which peaks at $\sim 5$ ($\sim 2$) for $H$-hadrons with exotic
 quarks (anti-quarks).


Within the
$H$-hadron kinematic region under study, the stopping of $H$-hadrons
is negligible, as can be understood from their low energy losses. The
decrease in speed ${\delta}\beta$ of a $H$-hadron following the
traversal of detector material in the ATLAS detector was calculated and it was found that
${\delta}\beta \lapprox 0.005$ (${\delta}\beta \lapprox 0.02$) for a mass of 200 GeV
(1000 GeV).

The results given above were obtained assuming that a neutral
$H$-hadrons could not oscillate into its anti-particle, something
which has been studied in the context of
supersymmetry~\cite{Gates:1999ei,Sarid:1999zx}. In a strictly
generic scenario of new physics, as considered here, it is prudent
to consider the possibility of two extreme cases: maximal and no
mixing. Allowing maximal mixing produces distributions (not shown),
which are very similar to those given in Fig.~\ref{fig_cont1}.
However, together with the effects of hadronisation and scattering
interactions it can give rise to striking event topologies, which
can be used to detect and characterise $H$-hadron, as is shown in
Section~\ref{rates}.

\subsection{Expected Rates of $H$-hadrons}\label{rates}

\begin{figure}[h!]
\begin{center}
\epsfig{file=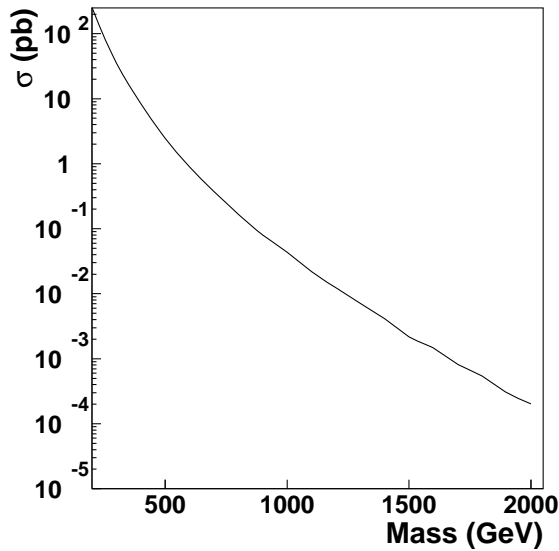,height=8cm,width=8cm} \caption{Predicted
cross section, calculated to leading order, for the pair production
of fourth generation quarks at the LHC.\label{fig_qu} }
\end{center}
\end{figure}

Fig.~\ref{fig_qu} shows the expected cross section for the pair
production of fourth generation quarks of a single flavour at the
LHC, as predicted by {\sc Pythia}. The  cross section is seen to
fall steeply with mass. Nevertheless, for a full year's LHC running
at low luminosity, corresponding to 10 fb$^{-1}$, around 25,000
pairs would still be expected for a 500 GeV mass, implying that
exotic hadrons can be sought with early LHC data.

Following the acceptance cuts and $H$-hadron transport through ATLAS
described in Section~\ref{eloss}, Tabs.~\ref{tab:1} and \ref{tab:2}
contain the expected yields for a range of event topologies arising
from the pair production of exotic quarks possessing $\pm
\frac{1}{3}e$ and $\pm \frac{2}{3}e$. Rates are shown for
$H$-hadrons of mass 200, 500 and 1000 GeV and an integrated
luminosity of 10~fb$^{-1}$ is assumed. The topologies are:

\begin{enumerate}
\item At
least one $H$-hadron is produced with non-zero charge and retains
that charge value throughout its passage in the detector material
implying that the inner and muon tracking chambers measure the same
charge;
\item Both particles are produced with non-zero charge and
retain those charge values.
\item Both particles are produced with
non-zero charge and one retains its non-zero charge but the other is
converted into a neutral state.
\item One particle is produced with
zero charge whilst the other has non-zero charge; the particle
produced with zero charge converts into a charged state whilst the
other retains its original value of charge.
\item Both particles are
produced with non-zero charge but leave the detector material as
neutral states.
\item At least one particle leaves the detector
material possessing a non-zero electric charge of opposite sign to
the charge with which it was produced.
\end{enumerate}

\begin{table}[h!]
\begin{center}
\begin{tabular}{|c|c|c|c|c|c|c|}
\hline
 &\multicolumn{3}{c|}{No mixing}&\multicolumn{3}{c|}{Maximal mixing}\\
Topology & \multicolumn{3}{c|}{Mass (GeV)} & \multicolumn{3}{c|}{Mass (GeV)}\\
\cline{2-7}
 & 200 &500 & 1000 & 200 & 500 & 1000 \\
\hline

1 & $4.9\times 10^5$   & $4.3\times 10^3$  & 57    & $4.1\times 10^5$    & $3.5\times 10^3$    & 48 \\
2 & $3.0\times 10^4$   & $2.6\times 10^2$  & 3    & $2.2\times 10^4$    & $1.9\times 10^2$    & 2  \\
3 & $9.6\times 10^4$   & $8.3\times 10^2$  & 9    & $8.2\times 10^4$    & $6.8\times 10^2$    & 8  \\
4 &  $6.0\times 10^4$  & $5.2\times 10^2$ & 6     & $4.8\times 10^4$    & $4.0\times 10^2$   & 5  \\
5 &  $6.4\times 10^4$  & $5.3\times 10^2$  & 6     & $6.3\times 10^4$    & $5.5\times 10^2$   & 6  \\
6 &   0                 & 0                 & 0    & $8.1\times 10^4$      & $7.2\times 10^2$  & 9  \\
 \hline
\end{tabular}\caption{Expected rates of various topologies, corresponding to an integrated luminosity of
10fb$^{-1}$, for different event topologies arising from the pair-production of exotic quarks of charge $\pm
\frac{1}{3}e$.}\label{tab:1}
\end{center}
\end{table}

\begin{table}[h!]
\begin{center}
\begin{tabular}{|c|c|c|c|c|c|c|}
\hline
 &\multicolumn{3}{c|}{No mixing}&\multicolumn{3}{c|}{Maximal mixing}\\
Topology & \multicolumn{3}{c|}{Mass (GeV)} & \multicolumn{3}{c|}{Mass (GeV)}\\
\cline{2-7}
 & 200 &500 & 1000 & 200 & 500 & 1000 \\
\hline

1 & $8.8\times 10^5$   & $8.0\times 10^3$     & 107    & $8.0\times 10^5$    & $7.2\times 10^3$    & 97  \\
2 & $1.2\times 10^5$   & $1.0\times 10^3$      & 12    & $8.6\times 10^4$    & $7.7\times 10^2$    & 8  \\
3 & $1.3\times 10^5$   & $1.2\times 10^3$  & 13    & $1.0\times 10^5$    & $8.9\times 10^2$    & 10  \\
4 &  $1.7\times 10^5$  & $1.5\times 10^3$ & 17     & $1.8\times 10^5$    & $1.5\times 10^3$   & 17  \\
5 &  $2.9\times 10^4$  & $2.5\times 10^2$  & 3     & $2.5\times 10^4$    & $2.1\times 10^2$   & 2  \\
6 &   0                 & 0       & 0    & $1.9\times 10^5$      & $1.7\times 10^3$  & 23  \\
 \hline
\end{tabular}\caption{Expected rates of various topologies, corresponding to an integrated luminosity of
10fb$^{-1}$, for different event topologies arising from the pair-production of exotic quarks of
charge $\pm \frac{2}{3}e$.}\label{tab:2}
\end{center}
\end{table}

 As can be seen, substantial rates for the mass points 200 and 500 GeV can be
  expected for each of the topologies. While the first two
  topologies offer the classic signature of a single or a pair of
  high momentum penetrating particles, scenarios 3 and 4 show
  evidence of charge exchange and can thus be used in combination with calorimeter
 information to identify the SMP as being hadronic and reject alternative scenarios,
  such as stable leptons. Scenario 5 may be
 the most experimentally challenging topology if a muon-based
 trigger is used, although, as seen, this topology accounts for only a
 small fraction of events. The observation of topology 6 could
 demonstrate mixing in the $H$-hadron sector.

A comparison of Tabs.~\ref{tab:1} and \ref{tab:2} shows the rates
for all of the considered topologies bar scenario 5 are greater for
$H$-hadrons with up-like exotic quarks than those for those with
down-like exotic quarks. However, the situation is reversed for
topologies (5). This can be explained as being due to the production
of baryons which have zero (non-zero) charge for $\pm \frac{1}{3}e$
($\pm \frac{2}{3}e$) charge exotic quarks.

\subsection{Detector Effects}
It is beyond the scope of this paper to study in detail the expected
response of the ATLAS detector. Such work is more appropriately made
with the ATLAS simulation and event reconstruction
packages~\cite{athena}. However, a few comments can be made based on
earlier work employing such programs and the results presented in
this Sections~\ref{eloss} and \ref{rates} should be considered in
the light of these.

It is desirable for any search to include calorimeter information. The observation of small hadronic energy
depositions associated with a penetrating SMP could be used, for example, to reject backgrounds of muons. Furthermore,
as shown in Fig~\ref{fig_cont1}, differences in hadronic energy loss between a pair of SMP candidates could provide
evidence for stable exotic quarks. To study the feasibility of exploiting calorimetry in this way, the
model presented here should be implemented within a package such as {\sc Geant} in order to include effects
such as nuclear fragmentation~\cite{geant}.

A conservative estimate of the trigger
efficiency for a slow moving particle using a muon trigger of 50\% has been
made~\cite{Kraan:2005ji}. Therefore, with the exception of topology 5, events corresponding to
the topologies listed above should be recorded in this way. It may be possible to
select events corresponding to topology 5, through a jet or missing transverse energy trigger, in the case when one of
exotic quarks emits a high transverse momentum quark or gluon.
The track reconstruction
efficiency in the inner and muon tracking chambers should be over 90\% in the kinematic region under
investigation~\cite{tdr}, though
further losses could occur due to charge exchange interactions of $H$-hadrons in the
different tracking systems. More studies on efficiency losses using, for example, the model
 presented in this work is desirable.
 Furthermore, the charge misidentification probability for
 tracks in the muon (inner) detector
should be around 2\% (4\%) for transverse momentum $p_T$ values up to around 1
TeV and falls quickly with decreasing $p_T$~\cite{tdr}. While it could be expected that the
 charge misidentification probability may be degraded for interacting $H$-hadrons, earlier work
 with gluinos involving a full ATLAS detector simulation showed that a signature
of an exotic hadron apparently reversing the sign of its charge is a useful search
observable~\cite{milstead-note}. We would thus expect topology 6 to be distinguished.


SM backgrounds can be suppressed in a number of ways, for example by making requirements
on the SMP's transverse momentum ($p_T \gapprox 100$ GeV) and event shape
 variables~\cite{Kraan:2005ji,milstead-note}. The most promising method is to use
a time-of-flight technique~\cite{Kraan:2005ji} with which exotic hadron searches in early LHC data
may only suffer from a handful of background events.



\section{Conclusions}
Heavy exotic quarks are predicted in a number of scenarios of
physics beyond the SM. This paper presented a generic model based on
Regge theory and the Quark Gluon String Model to describe the
interactions of heavy hadrons in matter. The work showed how strong
interactions of exotic hadrons may provide useful observable to
detect and quantify the properties of any exotic stable massive
particle which may be observed. Distributions relating to the
interactions and estimates of expected event topologies at the LHC
were shown. A natural next step in this work would be the
implementation of this model within ${\sc Geant}$ and its subsequent
use in searches at colliders and studies of discovery potential.

\section{Acknowledgements}
A. Kaidalov is partially supported by the following grants from the
Russian Foundation for Basic Research: RFBR 06-02-17012, RFBR
06-02-72041-MNTI, together with the Russian Scientific School Grant
843.2006.2.

D. Milstead is a Royal Swedish Academy Research Fellow supported by
a grant from the Knut and Alice Wallenberg Foundation.


\begin{thebibliography}{99}

\bibitem{Fairbairn:2006gg}
  M.~Fairbairn, A.~C.~Kraan, D.~A.~Milstead, T.~Sj\"ostrand, P.~Skands and T.~Sloan,
  Phys.\ Rept.\  {\bf 438} (2007) 1
  [arXiv:hep-ph/0611040].

\bibitem{ff}
  G.~R.~Farrar and P.~Fayet,
  Phys.\ Lett.\  B {\bf 76} (1978) 575
\bibitem{Gates:1999ei}
  S.~J.~Gates and O.~Lebedev,
  Phys.\ Lett.\  B {\bf 477} (2000) 216
  [arXiv:hep-ph/9912362].


\bibitem{s5d}
  M.~B.~Chertok, G.~D.~Kribs, Y.~Nomura, W.~Orejudos, B.~Schumm and S.~Su,
in {\it Proc. of the APS/DPF/DPB Summer Study on the Future of
Particle Physics (Snowmass 2001) } ed. N.~Graf, {\it In the
Proceedings of APS / DPF / DPB Summer Study on the Future of
Particle Physics (Snowmass 2001), Snowmass, Colorado, 30 Jun - 21
Jul 2001, pp P310}
  [arXiv:hep-ph/0112001].


\bibitem{sellis}
  J.~L.~Diaz-Cruz, J.~R.~Ellis, K.~A.~Olive and Y.~Santoso,
  JHEP {\bf 0705} (2007) 003
  [arXiv:hep-ph/0701229].

\bibitem{martin}
  S.~P.~Martin,
  Phys.\ Rev.\  D {\bf 75} (2007) 115005
  [arXiv:hep-ph/0703097].

\bibitem{collins}
P.D.B.~Collins, Phys.\ Rept.\  {\bf 1} (1971) 103. \\
P.D.B.~Collins, An Introduction to Regge Theory, {Cambridge (1977)}.


  \bibitem{QGSM}A.B.~Kaidalov, Phys.\ Lett.\  B {\bf 116} (1982) 459. \\
A.B.~Kaidalov and K.A.~Ter-Martirosyan, \Journal{\PLB}{117}{247}{1982} \\
 A.B.~Kaidalov and K.A.~Ter-Martirosyan, Sov.J.Nucl.Phys.{\bf 39}(1984)1545;\\
 A.B.~Kaidalov and O.I.~Piskounova, Sov.J.Nucl.Phys.{\bf 41}(1985)1278; \\
 O.~Piskounova, Phys.At.Nucl. {\bf 66} (2003) 332 [arXiv:hep-ph/0202005].


\bibitem{Ambrosanio:2000zu}
  S.~Ambrosanio, B.~Mele, A.~Nisati, S.~Petrarca, G.~Polesello, A.~Rimoldi and G.~Salvini,
  [arXiv:hep-ph/0012192].


\bibitem{Kraan:2005ji}
  A.~C.~Kraan, J.~B.~Hansen and P.~Nevski,
  Eur.\ Phys.\ J.\ C{\bf 49} (2007) 623
  [arXiv:hep-ex/0511014].

\bibitem{Johansen:2007kb}
  M.~Johansen,
  [arXiv:hep-ex/0701055].


\bibitem{Yao:2006px}
  W.~M.~Yao {\it et al.}  [Particle Data Group],
  J.\ Phys.\ G {\bf 33}, 1 (2006).


\bibitem{Baer:1998pg}
  H.~Baer, K.~m.~Cheung and J.~F.~Gunion,
  Phys.\ Rev.\  D {\bf 59} (1999) 075002
  [arXiv:hep-ph/9806361].
\bibitem{mod3} A.~Mafi and S.~Rabi, \Journal{\PRD}{62}{3}{1999}.

\bibitem{aafke_model}
A.C.~Kraan \Journal{\EPC}{37}{91}{2004}
\bibitem{Mackeprang:2006gx}
  R.~Mackeprang and A.~Rizzi,
  Eur.\ Phys.\ J.\  C {\bf 50}, 353 (2007)
  [arXiv:hep-ph/0612161].


\bibitem{geant}
{\sc Geant} - Detector description and simulation tool, CERN Program
Library Write-up, W5013, CERN, Geneva, 1993. \\
 S.~Agostinelli {\it et al.}  [GEANT4 Collaboration],
  Nucl.\ Instrum.\ Meth.\  A {\bf 506} (2003) 250.
 \bibitem{drees} M.~Drees and X.~Tata,
   \Journal{\PLB}{252}{695}{1990}.

\bibitem{kaidalov}A.B.Kaidalov,
Phys.Rep. {\bf 50}(1979)157.


\bibitem{Hooft} G.~t`Hooft, Nucl. Phys. {\bf B72} (1974) 461.
\bibitem{Veneziano} G.~Veneziano, Phys. Lett. {\bf B52} (1974)
220.

\bibitem{volkovitski}P.E.Volkovitski, A.B.Kaidalov, Sov. J. Nucl.Phys. 35(1982)1231.

\bibitem{Rossi} G.C.~Rossi and G.~Veneziano, Nucl. Phys. {\bf 45}
(1977) 507.


\bibitem{pythia}
T. Sj\"ostrand {\it et al.}, Comput. Phys. Commun. {\bf 135} (2001)
238.

\bibitem{string}
B.~Andersson, G.~Gustafsson, G.I.~Ingelman, and T.Sj\"ostrand, Phys.
Rep. {\bf 97} (1983) 31.



\bibitem{olga2}
  A.~B.~Kaidalov and O.~I.~Piskunova,
  Z.\ Phys.\  C {\bf 30} (1986) 145.




\bibitem{tdr} ATLAS Collaboration, {\it ATLAS: Detector and
Physics Performance Technical Design Report}, CERN-LHCC-99-14.


\bibitem{tile}
  P.~Adragna {\it et al.},
  IEEE Trans.\ Nucl.\ Sci.\  {\bf 53} (2006) 1275.


\bibitem{endc}
  D.~M.~Gingrich {\it et al.},
  JINST {\bf 2} (2007) P05005.


















\bibitem{Sarid:1999zx}
  U.~Sarid and S.~D.~Thomas,
  Phys.\ Rev.\ Lett.\  {\bf 85} (2000) 1178
  [arXiv:hep-ph/9909349].


\bibitem{athena}
  D.~Rousseau,
  Eur.\ Phys.\ J.\  C {\bf 33} (2004) S1038.




\bibitem{milstead-note} S.~Hellman, D.~Milstead and M.~Ramstedt,
ATLAS Public Note, ATL-PHYS-PUB-2006-005.





\end{thebibliography}
\end{document}